\documentclass[doublecol]{epl2}
\pdfoutput=1
\usepackage{amsmath,amssymb,xspace,graphicx,xcolor}
\usepackage{hyperref}
\usepackage{ulem}
\renewcommand{\r}{{\bf r}}
\newcommand{\n}{{\nabla}}
\newcommand{\p}{{\bf p}}
\newcommand{\D}{{\bf D}}
\newcommand\bfa{\boldsymbol \alpha}
\renewcommand{\u}{{\bf u}}

\newcommand{\J}{{\bf J}}
\newcommand{\Q}{{\bf Q}}
\newcommand{\V}{{\bf V}}

\newcommand{\red}[1]{{ #1}}

\newcommand\half{\frac{1}{2}}

\newcommand\calf{{\cal F}}

\newcommand\dO{{\rm d}\Omega}

\begin{document}

\title{\red{When are active Brownian particles and run-and-tumble particles equivalent? Consequences for motility-induced phase separation}}
\shorttitle{When are active Brownian particles and run-and-tumble particles equivalent? }
\author{M. E. Cates\inst{1}, J. Tailleur\inst{2}}
\institute{\inst{1}~SUPA, School of Physics, University of Edinburgh, JCMB
Kings Buildings, Edinburgh EH9 3JZ, United Kingdom\\ \inst{2}~Univ Paris Diderot, Sorbonne Paris \red{Cit\'e}, MSC, UMR 7057 CNRS, F75205 Paris, France}
\shortauthor{M. E. Cates and J. Tailleur}

\date{\today}

\abstract{Active Brownian particles (ABPs, such as self-phoretic
  colloids) swim at fixed speed $v$ along a body-axis ${\bf u}$ that
  rotates by slow angular diffusion. Run-and-tumble particles (RTPs,
  such as motile bacteria) swim with constant $\u$ until a random
  tumble event suddenly decorrelates the orientation. We show that
    when the motility parameters depend on density $\rho$ {but not
      on ${\bf u}$}, the coarse-grained \red{fluctuating
    hydrodynamics of interacting ABPs and RTPs can be mapped onto each
    other and are thus strictly equivalent.} In both cases, a steeply
  enough decreasing $v(\rho)$ causes phase separation in dimensions
  $d=2,3$, \red{even when no attractive forces act between the
  particles.} This points to a generic role for motility-induced phase
  separation in active matter. \red{However, we show that the
    ABP/RTP equivalence does not automatically extend to the
    more general case of $\u$-dependent motilities.}}

\maketitle

The physics of self-propelled colloidal particles represents a central
focus of research into `active matter' \cite{sriram,RoP,romanczuk}. In
active matter, a continuous supply of energy destroys microscopic
time-reversibility and allows phenomena to arise that are impossible
in thermal equilibrium systems, where detailed balance restores
time-reversal symmetry in the steady state.

One class of self propelled colloids is represented by motile bacteria
such as {\it Escherichia coli}, which move {in} a sequence of `runs'
-- periods of almost straight-line motion at near-constant speed $v$
-- punctuated by sudden and rapid reassignments of direction, or
`tumbles', occurring at random with rate $\alpha$
\cite{Berg,Schnitzer}. (Certain algae have similar motion
\cite{Goldstein}.) At large scales, and under conditions of constant
$v,\alpha$, a single run-and-tumble particle (RTP) performs a
random-walk of diffusivity $D_0 = v^2/\alpha d$ ($d$ is
dimensionality) which cannot be distinguished from the equilibrium
dynamics of non-interacting passive Brownian particles (PBPs).
Nonetheless, it was shown previously that if their swim speed
$v(\rho)$ decreases fast enough with the local particle density
$\rho$, RTPs undergo phase separation between a dense, slow-swimming
fluid and a dilute fast-swimming one \cite{TC}. This happens even in
the complete absence of interparticle \red{\textit{attractive}
  forces}, which are a prerequisite for fluid-fluid phase separation
in thermal PBPs. Coupled to population dynamics, this {\it
  motility-induced phase separation} was implicated in the patterning
of growing bacterial colonies \cite{PNAS,Hwa1,Hwa2}.

A second class of self-propelled colloids are represented by synthetic
(bi-)metallised colloids which, usually by catalysing the breakdown of
hydrogen peroxide, create a self-phoretic local chemical motor
\cite{Jones, PennState,Bocquet2010,golestanian,golestanian2}. These
differ from RTPs in that the direction of swimming only changes
gradually, by rotational diffusion (with angular diffusivity
$D_r$). This rotation is usually Brownian in origin, hence the name
ABPs (active Brownian particles). {Biological} ABPs also exist,
including non-tumbling {\it E coli} mutants and other organisms that
self-propel without tumbling. Clearly an isolated ABP again performs a
random walk ($D_0\propto v^2/D_r$) at large scales.

\red{When all microscopic parameters are uniform and isotropic, the
  large scale dynamics of RTPs, ABPs and PBPs are thus described by
  the diffusion equation. Equating their diffusivities thus trivially
  makes their dynamics identical. In the converse case, for
  instance when interactions, external potentials, or spatially
  varying swimming parameters are present, the dynamics become more
  complex and typically involve \red{a drift term:}
\begin{equation}
  \dot \rho = -\n\cdot [{\bf V}(v,\alpha,D_r,\dots) \rho- D(v,\alpha,D_r,\dots) \n \rho]
\end{equation}
While the microscopic parameters can clearly still be
  chosen so that the diffusivities of the three processes are equal,
  there is no reason why the equality of the drift terms ${\bf V}$
  should \red{then follow} as a by-product. Following
  Schnitzer~\cite{Schnitzer}, this point can be understood by
  considering PBPs in which either viscosity or temperature are
  non-uniform in space, \red{both leading to the same nonuniform
    $D({\bf r})$. As required by detailed balance for an isothermal
    system, ${\bf V}$ is zero in the first case despite the nonuniform
    diffusivity. But if the same diffusivity variation arises instead
    through a non-uniform $T({\bf r})$, detailed balance is broken,
    ${\bf V}$ is nonzero, and no generic Einstein-like relation
    relates $D$ to ${\bf V}$. Likewise, because ABPs and RTPs
    {are out of equilibrium}, the mere existence of a diffusive
    {behaviour}} in the uniform case does not establish \red{the}
  equivalence of ABPs with either PBPs or RTPs in more general
  situations. It is therefore intriguing that {recent
  simulations} of spherical ABPs with purely repulsive interactions
  showed clear signs of phase separation \cite{Fily,Redner}, something very
  similar to what is observed for RTPs~\cite{Alasdair}.}

By analogy with what we know for RTPs, such a phase separation is
explicable if the repulsions act to {effectively} decrease $v$ at high density (by
caging, for instance). However, {the growing} interest in ABPs
\cite{romanczuk} (exemplified by several other recent
theory/simulation studies and continuing experiments \cite{Dauchot,Peruani2,Ginelli,Stark1,Loewen1,Loewen2,Bocquet2012,SMF2012}) clearly requires a treatment of this
subject that lies beyond mere analogy. Are ABPs and RTPs actually
equivalent, and if so under what conditions?

This {Letter} provides a partial answer to that important question. At
time scales large compared to the angular reorientation time $\tau$
($\sim 1/D_r$ or $\sim 1/\alpha$), and length scales large compared to
$\ell = v\tau$, we {establish} a leading-order exact equivalence
between ABPs and RTPs so long as the motility parameters
$v,\alpha,D_r$ are independent of particle orientation ${\bf
  u}$. \red{This result, \red{which extends to} the case where a
  Brownian translational diffusivity $D_t$ is added, also holds when
  these parameters depend (slowly) on position. Following \cite{TC},
  this allows treatment of interacting particles, as long as the
  effect of interactions can be faithfully described by density-dependent swimming parameters. This can for instance be the case
when interactions are mediated by rapidly diffusing
chemicals~\cite{Hwa1,Hwa2} or when mean-field treatments are
valid (either as a result of molecular chaos or
when interactions are averaged over many neighbouring particles as
in~\cite{Alasdair}).} Thus the ABP-RTP equivalence indeed holds under
the conditions (generalised here to $d>1$) shown in \cite{TC} to cause
motility-induced bulk phase separation in RTPs. Accordingly ABPs {will
  generically show} the same sort of phase separation, subject to the
same requirement of a sufficiently decreasing $v(\rho)$.

Because we assume slow variations of all quantities on the scale of
$\ell$, we cannot rule out differences between ABPs and RTPs in
situations involving ordering at shorter scales, such as the
solidification transition addressed in \cite{Loewen1}. \red{Moreover,
  as we prove later on, no general equivalence exists for cases
  involving $\u$-dependent motility parameters. (This shows that while
  {\it some degree of similarity} between ABPs and RTPs might be
  {expected}, any {\it actual equivalence} between them is not
  automatic.)  Instances of such anisotropic dynamics include}
bacterial rectification \cite{Austin,Loewen2}, as modelled by an
orientation-dependent tumble rate \cite{TCEPL,RoP}, sedimentation,
where the particle speed acquires a gravitational component
\cite{TCEPL,Stark1,Bocquet2010} and systems of ABPs in which
interactions promote
alignment~\cite{Peruani1,Ginelli,romanczuk,gregoire,Mishra,McCandish,fred}.
Furthermore, hydrodynamic interactions, neglected entirely here, have
subtle consequences
\cite{Nash,Ishikawa,Pagonabarraga,Saintillan,Sokolov} that might well
differ between RTPs and ABPs. Despite these caveats, it is interesting
to establish an asymptotic equivalence between ABPs and RTPs in a
sector of parameter space that includes a nontrivial many-body
phenomenon, namely motility-induced bulk phase separation.

{\bf Fluctuating hydrodynamics}. We start by analysing a single
particle undergoing a generalized run-and-tumble dynamics with both
rotational and translational diffusions, and later recover RTP and ABP
as limiting cases.  We allow, $v$, $\alpha$, $D_r$ and $D_t$ to depend
on position but not orientation.  The probability density
$\psi(\r,\u,t)$ \red{of finding a particle at position $\r$ moving in
direction $\u$} obeys exactly (in $d=2,3$)
\begin{equation}\label{EoM1}
  \begin{aligned}
    &\dot\psi(\r,\u) = - \n.[v\u\psi(\r,\u)]+{\n_{\bf u} [D_r \n_{\bf u} \psi(\r,\u)]}\\
    &\quad+\nabla (D_t \nabla \psi(\r,\u))-\alpha \psi(\r,\u)
+\frac\alpha\Omega \int  \psi(\r,\u')\dO'
  \end{aligned}
\end{equation}
{where $\n_{\bf u}$ is the \red{rotational gradient} acting on
  ${\bf u}$} and the integral is over the unit sphere $|\bf u'|=1$ of
  area $\Omega$. The first term {on the right} is the divergence of
  the advective current resulting from self propulsion and the last
  two terms are loss and gain terms due to tumbling in and out of the
  direction ${\bf u}$. We next decompose $\psi$ as
\begin{equation}\label{decomp}
\psi(\r,\u,t) = \varphi + \p.\u +{\bf Q}:(\u\u-{\bf I}/d)+ \Theta[\psi]
\end{equation}
Here, $\varphi,\p,\Q$ are functions of $(\r,t)$ but not $\u$,
  which {parameterize} the zeroth, first and second angular ($d=2$) or
  spherical ($d=3$) harmonic components of $\psi$, while $\Theta$ projects
  onto the higher harmonic components~\cite{foot1}.

Integrating eq.~\eqref{EoM1} over $\u$ gives
\begin{equation}
\dot\varphi = -\frac 1 d\nabla(v {\bf p}) + \n (D_t \n \varphi) \label{EoM2}\\
\end{equation}
Using the orthogonality of angular/spherical harmonics, one can then
obtain equations order by order. Multiplying eq.~\eqref{EoM1} by {\bf u}
and integrating over $\u$ {then yields, after some
algebra~\cite{foot2},} for each component of ${\bf p}$
\begin{equation}
  \begin{aligned}
\label {EoM3}
\dot p_a &= -\n_a( v \varphi) - \big(D_r(d-1)+\alpha\big)
p_a + \n (D_t \n p_a)\\
&-B_{abcd}
\nabla_b (v Q_{cd})
  \end{aligned}
\end{equation}
where
$B_{abcd}=(\delta_{ac}\delta_{bd}+\delta_{ad}\delta_{bc}-
2 \delta_{ab} \delta_{cd}/d)/(d+2)$.

Similarly, multiplying eq.~\eqref{EoM1} by $\u \u -{\bf I}/d$ and
integrating over ${\bf u}$ yields {(with $\chi_{abc}$ defined below)}
\red{\begin{equation}
  \begin{aligned}
    \dot Q_{ab} &=  - \frac{d+2}{2} B_{abcd} \n_c (v p_d)-\n_c (\chi_{abc})\\
    &  - (2 d
    D_r+\alpha)   Q_{ab}+ \n (D_t \n  Q_{ab}) \label{EoM4}
  \end{aligned}
\end{equation}}

Equation~\eqref{EoM2} states that the rate of change of the probability
density of finding the particle at position $\r$ irrespectively of its
orientation (zeroth harmonic, $\varphi$) is the negative divergence of
a flux $\J = v\p/d-D_t\n \varphi$ which depends solely on the two
first harmonics. In addition to the diffusive flux $-D_t \n\varphi$,
there is a contribution of the self-propulsion along $\p$.

In eq.~\eqref{EoM3}, one sees that $\dot \p$ relaxes via a flux term
but also directly by angular diffusion and tumbling: here $(1-d)$ is
the eigenvalue of $\Delta_\u$ for the first harmonic (likewise {$-2d$} for
the second in eq.~\eqref{EoM4}). The last term gives the contribution
to $\dot \p$ arising from the $\Q$ term within $\n.[v\u\psi]$ in
eq.~\eqref{EoM1}. There are no contributions to $\dot \p$ from the
higher harmonics but there is one from the zeroth harmonic:
the $\n_a(v\varphi)$ term. It encodes the fact that, although
an isotropic distribution of swimmers cannot transport density if
$D_t=0$, it can still create anisotropy; for instance if the swim
speed or density is higher to the right of the origin, an initially
isotropic density at the origin soon develops an excess of left-moving
swimmers.

A corresponding set of remarks apply to eq.~\eqref{EoM4} for the
  time evolution of $\Q$. \red{Here $B_{abcd}\n_c(vp_d)$} is a flux contribution
  from the first harmonic whereas $\chi_{abc}$ (whose form we don't
  need) arises from the higher harmonics in {$\Theta[\psi]$}. Finally,
  the evolution of all remaining harmonics is obtained, if needed, by
  projecting eq.~\eqref{EoM1} using $\Theta$.

So far, beyond the assumed isotropy of $v(\r), D_{r,t}(\r)$ and
$\alpha(\r)$, no approximation has been made;
{eqs}.~(\ref{EoM2}-\ref{EoM4}) are exact results for the time
evolution of the zeroth, first and second harmonics of
$\psi(\r,\u,t)$. \red{At this stage, despite similar structures, the
  dynamics of ABP and RTP cannot be mapped onto each other. Equating
  the prefactors of $p_a$ in \eqref{EoM3} indeed requires
  $(d-1)D_r=\alpha$ which makes the prefactors of $Q_{ab}$ in
  \eqref{EoM4} unequal since then $2dD_r\neq \alpha$.}

\red{We} now
wish to coarse-grain these equations to obtain a `diffusion/drift'
expression for the flux $\J$ that involves only the conserved
probability density $\varphi(\r,t)$, the slowly varying parameters
$D_{r,t},v,\alpha$, and their first derivatives:
\begin{equation}\label{DDdef}
\dot\varphi = -\n.\J\;\;\;;\;\;\J = \V(\r,t)\varphi - D(\r,t)\nabla\varphi
\end{equation}
Following \cite{TC}, we use a gradient expansion and first note that
$\varphi$ is the only slow mode: its relaxation time is of order
$\sim (\n)^{-1}$ whereas all other harmonics relax in times of
order $\sim 1$. We thus assume $\dot\p=\dot\Q=\Theta[\dot\psi]=0$
when calculating the quasi-stationary current $\J$. By itself, this
creates a non-local constitutive relation between $\J$ and $\psi$
which still involves all harmonics. Next we carry out explicitly the
gradient expansion, yielding for $\Q$:
\red{
\begin{equation*}
Q_{ij} = -\frac{\frac{d+2}{2}B_{ijk\ell}\nabla_k (v p_\ell)+\nabla_k\chi_{ijk}}{2 d D_r +\alpha
}+{\cal O}(\n^2)
\end{equation*}}
from which the quasi-stationary $\p$ then follows via~\eqref{EoM3} as
\begin{equation} \label{pgrad}
\p = -\frac{1}{(d-1)D_r+\alpha} \n(v\varphi) + O(\n^2)
\end{equation}
{Fortunately therefore, to diffusion-drift order, closure is achieved
  without needing further information on harmonics beyond the first.}
To this order eq.~\eqref{DDdef} holds, with
\begin{equation}\label{DD}
\J = -\frac{v}{d(d-1)D_r+d\alpha}
\n(v\varphi)-D_t \n\varphi
\end{equation}
so that the diffusivity and drift velocity obey
\begin{equation}\label{VD}
D = \frac{v^2}{d(d-1)D_r+d\alpha}+D_t \;;\; \V = \frac{-v\n v}{d(d-1)D_r+d\alpha} 
\end{equation}

Eqs.~\eqref{DDdef} and \eqref{VD} give the evolution for the
probability density of one particle at diffusion-drift level. They are
equivalent to an Ito-Langevin equation for an individual particle
position $\r_\mu(t)$. Following \cite{TC} we can therefore now
consider an assembly of particles whose motility parameters
{$v,\alpha,D_r$ and $D_t$ depend on position through a set} of
(smooth) density functionals.  The coarse-grained density $\rho(\r,t)$
on which these depend then obeys the many body Langevin equation
\cite{TC}
\begin{equation} \label{coll}{
\dot\rho = -\n.\left(\V[\rho]\rho - D[\rho]\nabla \rho + (2D\rho)^{1/2}{\bf \Lambda}\right)
}
\end{equation}
with white noise $\langle \Lambda_i(\r,t)\Lambda_j(\r',t')\rangle =
\delta_{ij}\delta(\r-\r')\delta(t-t')$.  This noise term serves
  as a reminder that $\rho(\r,t)$ is not the one-particle probability
  $\varphi(\r,t)$ but a smooth coarse-graining of the collective
  density field $\sum_\mu\delta(\r -\r_\mu(t))$ (with $\mu$ a particle
  index) which evolves stochastically. The functionals $v[\rho]$,
  $\alpha[\rho]$ and $D_{t,r}[\rho]$ in~\eqref{VD} then defines for
the interacting particle system the many-body drift velocity and
diffusivity $\V[\rho]$ and $D[\rho]$ for use in eq.~\eqref{coll}.

We can now use eqs.~(\ref{VD},\ref{coll}) to compare the dynamics of
RTPs and ABPs. From the expression of $D$ and $\V$ in~\eqref{VD}, we
see that the tumble rate $\alpha$ and rotational diffusivity $D_t$
enter only through the combination $(d-1)D_r+\alpha \equiv \tilde
\alpha$. {Hence} the pure RTP and pure ABP limits, alongside
anything in between, are made equivalent by a suitable choice of
$\tilde\alpha[\rho]$. The effect of nonzero $D_t$ is the same in each
case, although we note that, in many experimental cases, the
self-propelling speed $v$ is large enough that the translational
diffusivity $D_t$ is negligible. (For instance in wild-type
run-and-tumble bacteria, $v^2/(d\alpha)$ is about three orders of
magnitude larger than $D_t$.) In this limit, pure ABPs
($D_t=\alpha=0$) and pure RTPs ($D_t=D_r=0$) are equivalent up to the
mapping $(d-1)D_r\leftrightarrow\alpha$.

{\bf Motility-induced phase separation.} The many-body physics
predicted by eqs.~(\ref{VD},\ref{coll}) for {the large class of active
  particle systems addressed above essentially coincides with that
  studied in~\cite{TC} for RTPs in $d=1$. (Note that ABPs cannot be
  defined in $d=1$.)} As shown there, because~\eqref{coll} is
equivalent to a set of interacting PBPs with the specified diffusivity
and drift functionals, a condition can be found under which the
interacting active system will behave just like a fluid of PBPs with
free energy (in thermal units) $\calf[\rho] = \calf_{ex}[\rho] +
\int\rho(\ln\rho-1)dx$.  The required condition is \cite{TC}
\begin{equation}
\V([\rho],\r)/D([\rho],\r) = -\nabla(\delta \calf_{ex}[\rho]/\delta \rho(\r)) \label{integrable}
\end{equation} 
where the right hand side represents the force ({\it i.e.}, excess
chemical potential gradient) on a particle at $\r$. 

In the case where $D_t=0$, the left hand side of~\eqref{integrable} is
simply $-\nabla \ln v[\rho]$ and we then require $\delta
\calf_{ex}[\rho]/\delta\rho(\r) = \ln(v([\rho];\r)$. When the swim
speed depends only locally on density, so that $v([\rho];\r) =
v(\rho(\r))$, we have $\calf_{ex} = \int f_{ex}(\rho(\r)) d\r$ where
$f_{ex} = \int_0^\rho\ln v(s)ds$. (Note that when $\tilde\alpha
  d {D}_t $ is a nonzero constant, this result generalizes to $f_{ex}
  = \half\int_0^\rho\ln[ v(s)^2+\tilde\alpha d {D}_t] ds$.) This form of excess free energy leads at
mean-field level to a spinodal instability whenever {$dv/d\rho <
  -v/\rho$}, and to binodal conditions for phase coexistence given by
the usual common tangent construction on $f = f_{ex} +
\rho(\ln\rho-1)$ \cite{TC}. Beyond mean field, the steady state
probability for density fluctuations is governed by the
`Boltzmann-like' distribution $\exp[-\calf]$ and these fluctuations
will eventually cause phase separation at all densities between the
binodals.

To understand the phase-separation dynamics in detail, to address the
interfacial tension between phases, and also to confirm that the phase
separation is equilibrium-like (rather than having, say, an
ever-moving interface between the phases), would require a detailed
examination of gradient terms lying beyond the present study. The same
was initially true of the analysis of RTPs made in \cite{TC}, but the
relevant gradient terms were later identified explicitly
\cite{Alasdair}. However, some of the gradient terms neglected via
eq.~\eqref{pgrad} could in principle differ between ABPs and RTPs,
even if these share the same functionals $\V[\rho]$ and
$D[\rho]$. Accordingly, the somewhat technical exploration of these
higher gradient terms is, for the ABP case, deferred to future work.

\begin{figure}
  \includegraphics[width=.4\columnwidth]{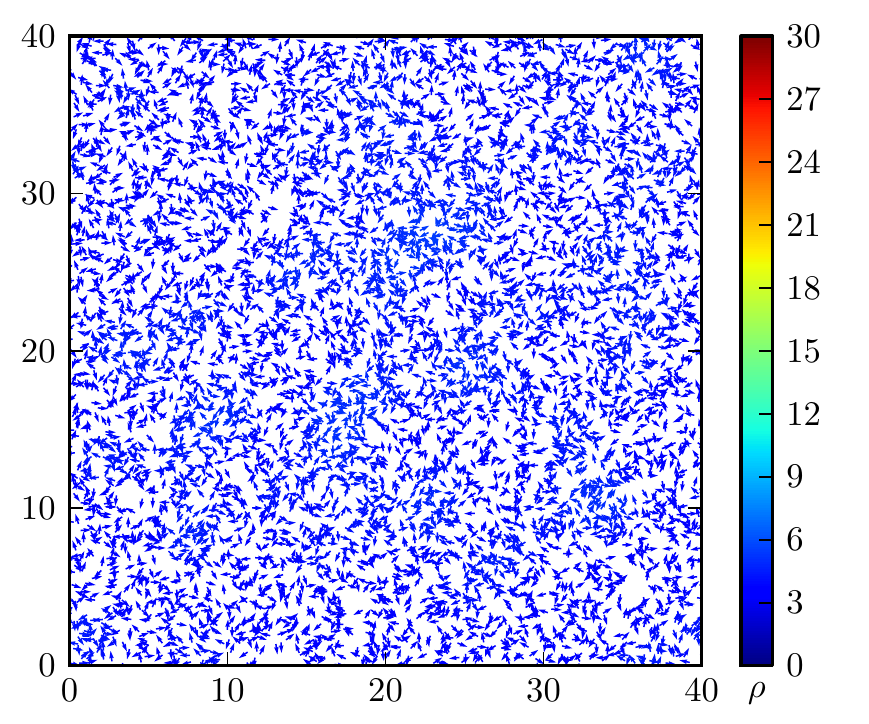} \hspace{.1\columnwidth}  \includegraphics[width=.4\columnwidth]{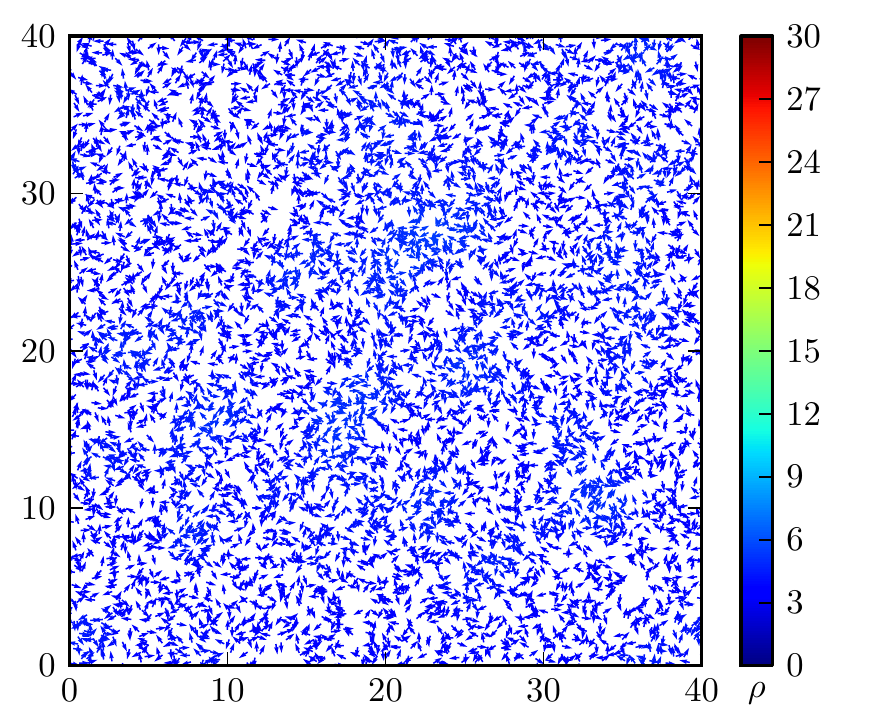}\\
  \includegraphics[width=.4\columnwidth]{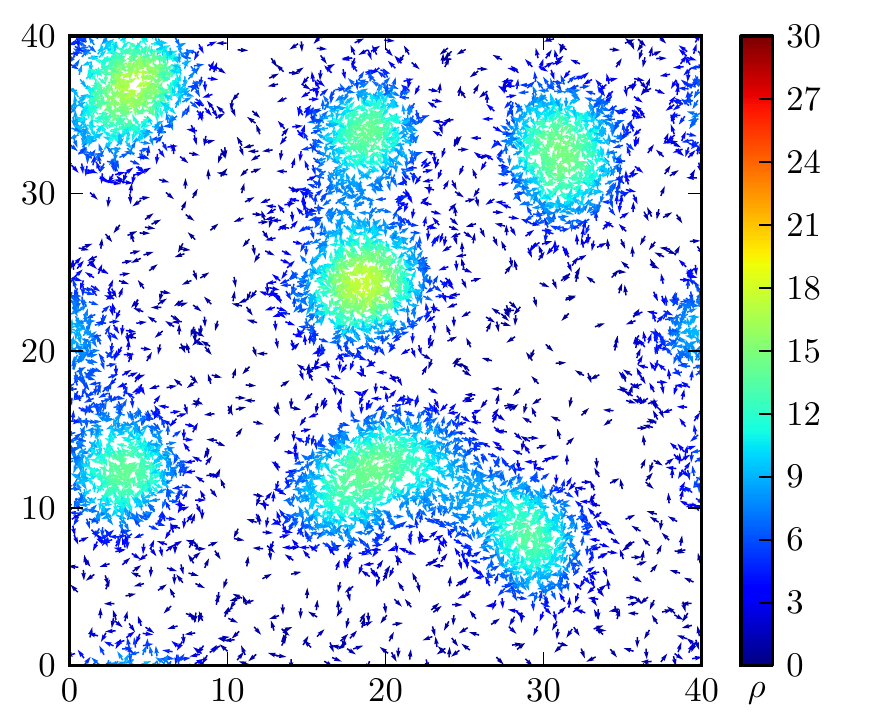} \hspace{.1\columnwidth} \includegraphics[width=.4\columnwidth]{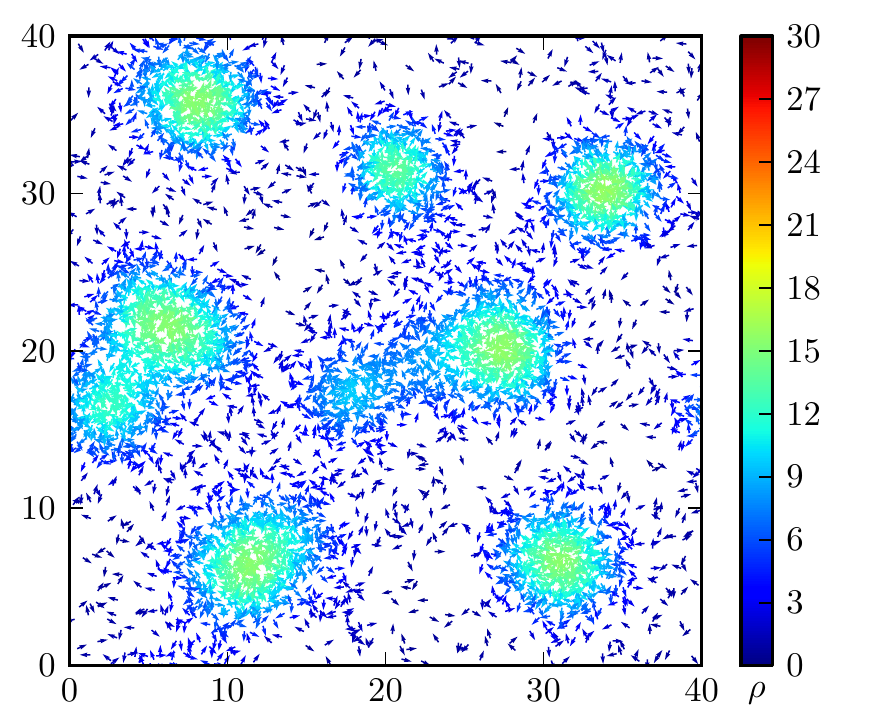}\\
  \includegraphics[width=.4\columnwidth]{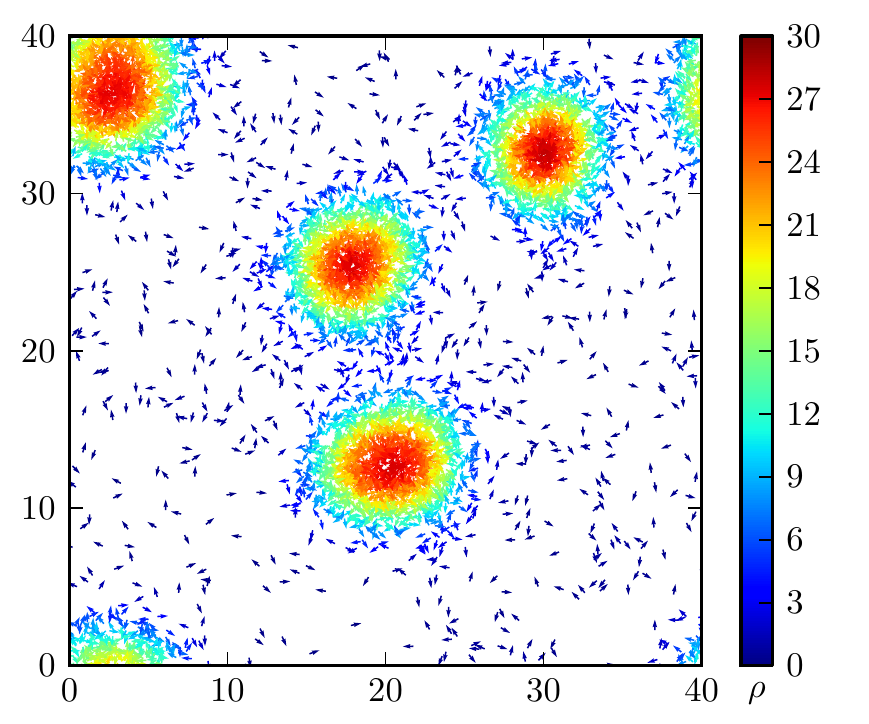}\hspace{.1\columnwidth}  \includegraphics[width=.4\columnwidth]{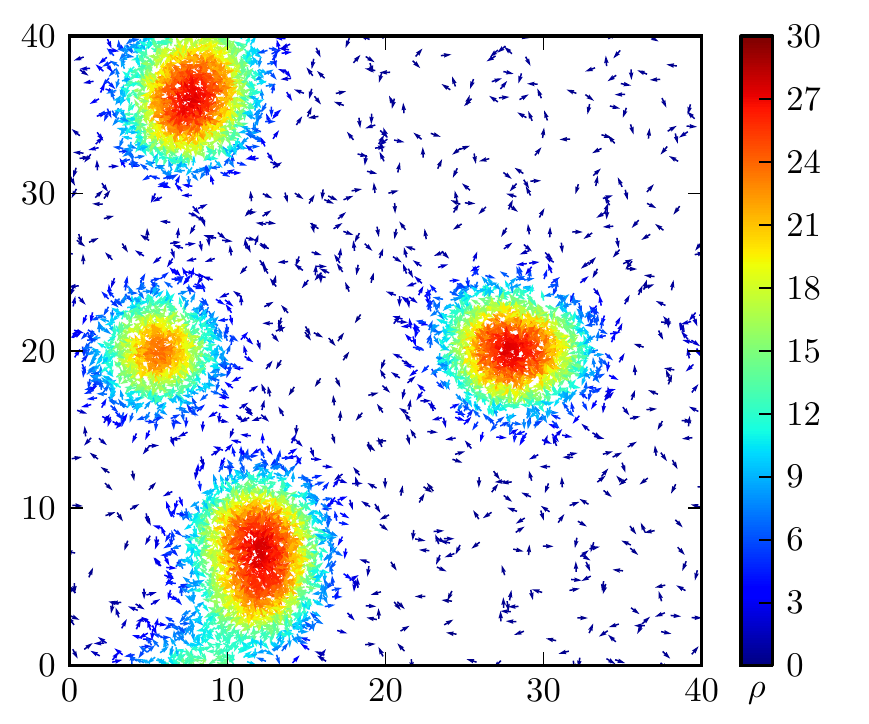}\\
  \caption{Spinodal decomposition of $6400$ ABP (left) and RTP (right)
    in 2D with $v_0=10$, $\alpha=D_r=1.25$, $\lambda=0.45$,
    $\phi=7$, $w=3.5$ at times 0 (top), 100 (middle) and 1000 (bottom). Each
    particle appears as an arrow pointing along its body axis and with
    a colorcode equal to the local density $\rho(\r)$.}
  \label{fig:spin}
\end{figure}

The similarity between RTPs and ABPs undergoing phase separation can
nevertheless be explored numerically {by comparing simulations of
  large populations of interacting ABPs and RTPs}. To do so we {set
  $D_t = 0$ and} consider a swimming speed
$v[\rho(x)]=v_0\exp[-\lambda \phi \arctan(\rho/\phi)]$ which decreases
exponentially at low density before saturating at finite but non-zero
swimming speed\footnote{\red{As in~\cite{TC}, the density in the
    dense phase diverges in the absence of excluded volume
    interactions if $v(\rho)$ vanishes at large $\rho$.}} for
larger $\rho$. Numerically, $\rho(\r)$ is computed by convoluting the
number density $\sum_\mu \delta(\r-\r_\mu)$ with the function $f(\r)=Z
\exp[-1/(1-r^2/w^2)]$ with $Z$ a normalisation constant. For such a
$v[\rho(x)]$, the free energy is bimodal as soon as $\lambda
\phi>2$. Starting from identical uniform initial condition, the
spinodal decomposition of $6400$ RTPs and ABPs is shown in
figure~\ref{fig:spin} and is consistent with the aforementioned
mapping between ABPs and RTPs.

We can now turn back to the simulations {of \cite{Fily,Redner}} which show
that excluded volume interactions promote phase separation of
ABPs. This result, {like that found numerically} for RTPs on a
lattice~\cite{Alasdair}, fits well with the scenario presented here,
even though the interactions in~\cite{Fily,Redner} are collisional rather
than via a smooth density functional. Notably, the collisions
considered in~\cite{Fily,Redner} do not directly affect the orientation $\u$;
at high density a particle undergoes many collisions before one-body
rotational diffusion relaxes its swimming direction. Modelling
collisions via $v(\rho)$ seems reasonable in this case. Indeed,
$v(\rho)$ was reported in~\cite{Fily} to decay linearly, a case shown
in~\cite{Alasdair} to cause phase separation for RTPs. Although a
microscopic model starting directly with an effective $v(\rho)$ is
only a mean-field-like description of the ABP simulations
of~\cite{Fily,Redner}, the mechanistic similarity between their phase
  separation and that of RTPs is now clear.

  \red{{\bf Anisotropic motilities.} We now turn to more general cases where
  the \red{dynamics} is not isotropic and discuss the limits of the
  equivalence between ABPs and RTPs. There are many ways of breaking
  isotropy and, as we show below, the mapping between RTPs and ABPs
  can be violated qualitatively, quantitatively, or preserved, depending
  on details of the microscopic dynamics. 

One simple way to remove isotropy is to add a uniform field (such as
gravity) that superimposes a Stokes drift velocity ${\bf w}$ on the
self-propulsion. The overall velocity of a particle is then $v \u+{\bf
  w}$. This only affects the advection term in~\eqref{EoM1} and adds
$- \n({\bf w}\varphi)$, $- \n_b({w_{b}}p_a)$ and $-
\n_c({w_{c}}Q_{ab})$ to the right hand side of
  equations~(\ref{EoM2}-\ref{EoM4}). The large time/scale expansion
  then yields $Q\sim\n$ as before and one gets for $p$ and $\varphi$
\begin{equation}
  \begin{aligned}
    \label{eqn:extpot}
    \dot \varphi&=-\frac{1}{d}\n({v} \p) - \n({\bf w}\varphi)+ \n (D_t \n \varphi)\\
\big(D_r(d-1)+\alpha\big)
    p_a &= -\n_a( v \varphi)-\n_b( w_{b} p_a) +{\cal O}(\n^2)
  \end{aligned}
\end{equation}
At this coarse-grained level, there is again an exact equivalence
between RTPs and ABPs upon identifying $\alpha$ and $ (d-1) D_r$. This
is particularly interesting since a uniform field can induce polar
order~\cite{Stark1}, which thus does not by itself breaks the mapping
between ABPs and RTPs.

Another way of breaking isotropy is to have the swimming parameters
depend on the local orientation $\u$. For simplicity, we will
only consider a first order spherical harmonic dependence
$D_r(\u)=D^r_0+\D^r_1\cdot\u$ and
$\alpha(\u)=\alpha_0(\u)+\bfa_1\cdot \u$. We will also take
$D_t=0$ for sake of clarity. Equation~\eqref{EoM1} now becomes
\begin{equation}\label{psidotstrato}
  \begin{aligned}
  \dot\psi &= - \n.[v\u\psi]+\n_\u \cdot [(D^r_0 +\D^r_1 \cdot \u) \n_\u \psi]\\
  &-(\alpha_0+\bfa_1\cdot \u) \psi + \alpha_0 \varphi +\frac 1 d \bfa_1 \cdot \p \\
  \end{aligned}
\end{equation}
Once again, we project this equation onto the three first
harmonics. Tedious but straightforward algebra, that will be detailed
elsewhere, show that equation~\eqref{EoM2} is not
modified whereas~(\ref{EoM3},\ref{EoM4}) become
\begin{equation}\label{eq:pstrato}
  \begin{aligned}
  \dot p_i =& -\n_i(v\varphi)-[\alpha_0+(d-1)D_0^r] p_i -\alpha_{1,i} \varphi\\& -B_{ijkl} [\n_j(v Q_{kl})+\alpha_{1,j}Q_{kl} +d D^r_{1,j}Q_{kl}]
  \end{aligned}
\end{equation}
and \red{
\begin{equation*}
  \begin{aligned}
    \frac{2 \dot Q_{ij}}{d+2} =&
    -B_{ijk\ell}\n_k[vp_\ell]-\frac{d}{\Omega} \int \dO S^{(3)}_{ij k}
    S^{(3)}_{lmn } \n_k[v T_{lmn}]\\&-\frac{2\alpha_0+ 4d
      D_0}{d+2}Q_{ij}-B_{ijk\ell}(\alpha_{1,k}+ d D_{1,k})p_\ell\\&
    -\frac{d}{\Omega} [\alpha_{1,k}+2 (d+1) D_{1,k}]T_{\ell m n}\int
    \dO S^{(3)}_{i j k}S^{(3)}_{\ell m n}
  \end{aligned}
\end{equation*}}
where $S^{(3)}_{ijk}=u_i u_j u_k-(u_i\delta_{jk}+u_j
\delta_{ik}+u_k\delta_{ij})/(d+2)$ is a tensor of 3rd order spherical
harmonics and $T_{ijk}$ the component of $\psi$ along this
tensor.

Equation~\eqref{eq:pstrato} already shows that whatever the choice of
coefficients, there can never be an equivalence between ABPs and RTPs
in this context, since the tumbling yields a contribution
$-\alpha_{1,i}\varphi$, which has no counterpart stemming from
the rotational diffusion and which does not disappear in the large
time/scale limit. Interestingly, the addition of a well chose angular
potential for the rotational diffusion partially solves this
problem. Such a potential produces an angular drift term
  $\dot\psi \sim - \n_\u({\bf V}_\u\psi)$; choosing a drift ${\bf
    V}_\u = -\n_\u D_r$ then effectively replaces the term $\n_\u
  (D(\u) \n_\u \psi)$ in ~\eqref{EoM2} by $\Delta_\u (D(\u) \psi)$,
  which adds a $-(d-1)D_{1,i}\varphi$ term to~\eqref{eq:pstrato}. If
  this is done, the hydrodynamic equations for ABPs and RTPs do have
  similar structures, but it is still not possible to choose motility
  parameters to make all the coefficients match. Thus the mapping
  breaks down quantitatively. While these technical subtleties will be
  presented in more detail elsewhere, the results above are enough to
  prove that no general equivalence between ABPs and RTPs exists, once
  anisotropic dynamics is present.}

{\bf Discussion.} We comment finally on the role of the effective free
energy \eqref{integrable} in relation to {the concept of} effective
temperature. Several works have shown that active baths can be
described by effective temperatures when dealing with passive
tracers~\cite{Wu,Leticia1,Leticia2} or external
potentials~\cite{TCEPL,Bocquet2010}, {which implies that active
  particles may be considered as \red{'hot colloids'}}. For interacting ABPs,
however, the usefulness of such effective temperatures has been
questioned~\cite{Loewen1,Fily}. We have shown above that the
phenomenology expected for interacting ABPs is fundamentally different
from that of PBPs at any temperature. Nevertheless, under the specific
conditions addressed here for which \eqref{integrable} is obeyed,
equilibrium-like concepts do apply: the steady states of ABPs then
satisfy detailed balance and so admit a free energy
description. Currently we do not know the conditions (if any) under
which a path can be traced from the microscopic dynamics of
interacting ABPs directly to an effective free energy, avoiding the
coarse graining steps that led us from \eqref{EoM1} to
\eqref{integrable}. More generally, the validity or otherwise of
effective temperature descriptions may largely depend on whether
detailed balance emerges in a coarse grained description even though
it is, by definition, absent microscopically in all active~systems.

{\bf Conclusion.} In summary, this {Letter establishes} that, under
conditions where their local swim speeds and reorientation rates do
not depend on the swimming direction, a large class of active particle
{models obey a common} equation at the diffusion-drift level. {The ABP
  and RTP limits are connected by a mapping, $\alpha \leftrightarrow
  (d-1)D_r$,} between tumble rate $\alpha$ and rotational diffusivity
$D_r$. Were tumbles to be considered of finite duration, one would
expect a more complex mapping in which the swim speed $v$ of the
equivalent ABPs also depends on $\alpha$ \cite{TC}.

The diffusion-drift description treats all nonconserved density
harmonics as quasi-stationary, and expands the total particle flux to
first order in spatial gradients; it therefore requires slow
variations in the conserved density (or zeroth harmonic). Since a
description at this level is sufficient to explain the
motility-induced phase separation of RTPs whose swim speed is
density-dependent, ABPs will show such phase separations under exactly
the same conditions on $v(\rho)$ as derived previously for RTPs. (In
passing we have clarified that those conditions apply in $d=2,3$ as
well as for $d=1$ as derived originally \cite{TC}.) Motility-induced
phase separation is caused by a feedback between (a) the tendency for
particles to move slowly in regions of high density and (b) the
tendency for particles to accumulate where they move slowly. The
second of these is prohibited by detailed balance in thermal
equilibrium. {Our work shows that such phase separation (derived previously only for RTPs in 1D) is a robustly generic feature of
interacting active particles.}

\red{Furthermore, in cases where the \red{dynamics} is not isotropic, we have
  shown that the mapping can \red{sometimes survive (even in the presence of polar
  order)} but can also \red{break down} because of new couplings between
  harmonics of different orders. Nonetheless, progress might be
  possible when the angular dependencies on $\u$ are in turn
  expandable in low order angular or spherical harmonics as was the
  case in~\cite{fred}. 

\red{Although the approach presented in this letter 
  already can} treat the sedimentation {\cite{Bocquet2010,Stark1}}
  and rectification \cite{Austin} problems alluded to previously,
  direct interparticle forces \red{remain} to be explored.} For example,
interparticle attractions~\cite{PoonPNAS,Bocquet2012} or nematic
interactions~\cite{Peruani1,Ginelli} promote clustering tendencies
that seem distinct from the bulk phase separation caused by motility
modulations alone. In general, interparticle forces create local
orientational variation of the particle speed (just as the uniform
force of sedimentation creates a global variation) \cite{TC} which
effectively means $v = v(\u)$ or $v = v(\p)$. Accordingly, despite
some initial steps that have been made recently
\cite{gregoire,Mishra,Stark1,fred} an explicit treatment of that case
appears necessary as a first step towards a general dynamic density
functional theory for interacting ABPs. Such a theory, which might
involve analysis connected with that done for liquid crystals in
\cite{Raphael}, is a worthy goal for future research.

\begin{acknowledgments} The authors thank Suzanne Fielding, Alasdair
Thompson and Raphael Wittkowski for discussions. MEC thanks the Royal
Society for a Research Professorship and KITP Santa Barbara for
hospitality. JT thanks the Weizmann Institute of Science and the
Technion Physics Department for hospitality. This work was supported
in part by EPSRC Grant EP/J007404/1 and in part by the National
Science Foundation under Grant No. NSF PHY11-25915.
\end{acknowledgments}


\begin{thebibliography}{99}
\bibitem{foot2} The derivation of Eq.~(\ref{EoM2}-\ref{EoM4}) rely on
  the fact that spherical and angular harmonics of order $\ell$ are
  eigenvectors of $\Delta_\u$ with eigenvalues $-\ell (\ell + 1)$ and
  $ -\ell^2$ , respectively and on the equalities $\int u_\alpha
  u_\beta=\frac{\Omega}d \delta_{\alpha\beta}$, $\int u_\alpha^4
  \dO=\frac{3\Omega}{d (d+2)}$ and $\int u_\alpha^2
  u_{\beta\neq\alpha}^2 \dO=\frac{\Omega}{d (d+2)}$ where $\Omega$ is
  the total solid angle of the unit sphere.
\end{thebibliography}
\end{document}